\documentclass[a4paper, 11pt]{article}
\usepackage{amssymb}
\usepackage{amsmath}

\newcommand{\be}{\begin{equation}}
\newcommand{\ee}{\end{equation}}
\newcommand{\bea}{\begin{eqnarray}}
\newcommand{\eea}{\end{eqnarray}}

\newcommand{\ri}{\mathrm{i}}

\newcommand{\bN}{\mathbb{N}}
\newcommand{\bR}{\mathbb{R}}
\newcommand{\bC}{\mathbb{C}}

\newcommand{\bZ}{\mathbb{Z}}

\newcommand{\cB}{\mathcal{B}}
\newcommand{\cC}{\mathcal{C}}
\newcommand{\cD}{\mathcal{D}}

\newcommand{\cL}{\mathcal{L}}
\newcommand{\cP}{\mathcal{P}}
\newcommand{\cR}{\mathcal{R}}

\newcommand{\mfa}{\mathfrak{a}}
\newcommand{\mfc}{\mathfrak{c}}
\newcommand{\mfm}{\mathfrak{m}}
\newcommand{\mfu}{\mathfrak{u}}
\newcommand{\mfgl}{\mathfrak{gl}}

\newcommand{\1}{\boldsymbol{1}}
\newcommand{\bsq}{\boldsymbol{q}}
\newcommand{\bslambda}{\boldsymbol{\lambda}}
\newcommand{\bschi}{\boldsymbol{\chi}}

\newcommand{\diag}{\mathrm{diag}}

\newcommand{\ad}{\mathrm{ad}}
\newcommand{\wad}{\widetilde{\ad}}

\newcommand{\tr}{\mathrm{tr}}
\newcommand{\Id}{\mathrm{Id}}

\newcommand{\reg}{\mathrm{reg}}
\newcommand{\sgn}{\mathrm{sgn}}

\newcommand{\eps}{\varepsilon}

\newcommand{\qedsymb}{$\blacksquare$}
\begin{document}
\begin{center}
\Large{\textbf{On the scattering theory
of the classical hyperbolic $C_n$ Sutherland model}}
\end{center}
\bigskip
\begin{center}
B.G.~Pusztai\\
Bolyai Institute, University of Szeged,\\
Aradi v\'ertan\'uk tere 1, H-6720 Szeged, Hungary\\
e-mail: \texttt{gpusztai@math.u-szeged.hu}
\end{center}
\bigskip
%%%%%%%%%%%%%%%%%%% ABSTRACT %%%%%%%%%%%%%%%%%%%%%%
\begin{abstract}
In this paper we study the scattering theory of the
classical hyperbolic Sutherland model associated with
the $C_n$ root system. We prove that for \emph{any} 
values of the coupling constants the scattering map 
has a factorized form. As a byproduct of our analysis,
we propose a Lax matrix for the rational $C_n$ 
Ruijsenaars--Schneider--van Diejen model with
\emph{two independent coupling constants}, thereby 
setting the stage to establish the duality between the
hyperbolic $C_n$ Sutherland and the 
rational $C_n$ Ruijsenaars--Schneider--van Diejen models.
\end{abstract}
\newpage
%%%%%%%%%%%%%%%%%%%%%%%%%%%%%%%%%%%%%%%%%%%%%%%%
%%% section: Introduction
%%%%%%%%%%%%%%%%%%%%%%%%%%%%%%%%%%%%%%%%%%%%%%%%
\section{Introduction}
In the study of interacting many-particle systems 
it is hard to overestimate the importance of scattering
theory. At the same time, it is notoriously difficult 
to obtain rigorous results in this subject. It is a 
very fortunate situation that for certain integrable 
systems, defined on the real line, the scattering 
theory has been completely understood. In particular, 
we have full control over the scattering theory of 
the Toda systems, the Calogero--Moser--Sutherland models, 
and the Ruijsenaars--Schneider models associated with 
the $A_n$ root system (see e.g. \cite{Moser1975}, 
\cite{Moser1977}, \cite{Kulish1976}, \cite{RuijCMP1988}, 
\cite{RuijFiniteDimSolitonSystems}).
Their characteristic feature is that the scattering 
map has a factorized form. 
Though these integrable many-particle systems
have natural generalizations to other root systems
as well (see e.g. \cite{OlshaPere}), 
the scattering theory of the non-$A_n$-type models 
is far less developed than that of the $A_n$-type 
systems.

In this paper we undertake the task to understand the
scattering behavior of the classical hyperbolic $C_n$ 
Sutherland model. (For background information on the
$C_n$-type Sutherland systems see e.g.
\cite{OlshaPere}, \cite{AvanBabelonTalon1994},
\cite{FeherPusztai2006}, \cite{FeherPusztai2007}.) 
Recall that the phase space of this model is the 
cotangent bundle of the Weyl chamber 
$\mfc := \{ q = (q_1, \ldots, q_n) \in \bR^n 
\, | \,  
q_1 > \ldots > q_n > 0 \}$ and the dynamics
is governed by the Hamiltonian
\be
H_{C_n}(q, p) 
= \frac{1}{2} \sum_{c = 1}^n p_c^2
+ \sum_{1 \leq a < b \leq n}
\left(
\frac{g^2}{\sinh^2(q_a - q_b)}
+ \frac{g^2}{\sinh^2(q_a + q_b)}
\right)
+ \sum_{c = 1}^n \frac{g_2^2}{\sinh^2(2 q_c)},
\label{H}
\ee
where $g$ and $g_2$ are arbitrary non-zero real
numbers, the so-called coupling parameters.
By the repulsive nature of the interaction we 
expect that the particles move asymptotically
freely for very large positive and negative values 
of time $t$, thus it makes sense to study the
scattering map that relates the asymptotic phases
and momenta of the past and the future. 
Using only elementary algebraic techniques, 
the main goal of the paper is to show that the 
scattering map of the $C_n$-type model also has 
a \emph{factorized form}, i.e., the classical phase 
shifts are entirely determined by the two-particle 
processes and by the one-particle scatterings on 
the external field. The precise statement is
given in Theorem 3. Though this result does meet 
our expectations, to our knowledge, its rigorous 
proof has not appeared in the literature before.

To understand the scattering properties of the hyperbolic 
$C_n$ Sutherland model we closely follow Ruijsenaars' 
seminal work \cite{RuijCMP1988} on the $A_n$ system.
One of the upshots of his approach is that it reveals a 
natural \emph{action-angle duality} between the hyperbolic 
Sutherland and the rational Ruijsenaars--Schneider 
models. Surprisingly, for root systems other than $A_n$,
relatively few is known about the duality between the
Sutherland and the Ruijsenaars--Schneider--van Diejen
(RSvD) models. Even the Lax representation of the generic 
RSvD dynamics is missing, except for some very special 
one-parameter family of $C_n$ and $BC_n$ models obtained 
by the natural $\bZ_2$-folding of the original 
$A_{2 n - 1}$ and $A_{2 n}$  systems \cite{ChenHouYang0006}. 
Only partial results \cite{ChenHou0102} are known for the 
$D_n$ root system, too. However, as a byproduct of our 
scattering theoretic analysis, we obtain a natural 
candidate for the \emph{Lax matrix} of the two-parameter 
family of rational $C_n$ RSvD models. The Lax matrix 
presented in Lemma 2 generalizes the known 
one-parameter family of rational Lax matrices 
obtained by the folding procedure and it generates 
the rational $C_n$ RSvD Hamiltonian \cite{vanDiejen1994} 
with two independent coupling parameters. 

The paper is organized as follows. To keep the 
presentation self-contained, in Section 2 we 
collect the necessary background material and 
fix the notational conventions. In Section 3 
we present our results on the scattering
theory of the hyperbolic $C_n$ Sutherland model. 
After the discussion part in Section 4, we finish 
the paper with an Appendix on some useful facts 
from linear algebra. 
%%%%%%%%%%%%%%%%%%%%%%%%%%%%%%%%%%%%%%%%%%%%%%%%%%%%%%%
%%% SECTION: PRELIMINARIES
%%%%%%%%%%%%%%%%%%%%%%%%%%%%%%%%%%%%%%%%%%%%%%%%%%%%%%%
\section{Preliminaries}
In this section we gather some basic facts about 
the hyperbolic $C_n$ Sutherland model. We start
with a short review on some group theoretic material 
related to the non-compact Lie group $U(n, n)$, then 
we discuss the Lax representation of the Sutherland 
dynamics. For the details the reader may consult 
\cite{Helgason}, \cite{Knapp}, \cite{OlshaPere}, 
\cite{AvanBabelonTalon1994}, \cite{FeherPusztai2006}, 
\cite{FeherPusztai2007}.
%%%%%%%%%%%%%%%%%%%%%%%%%%%%%%%%%%%%%%%%%%%%%%%%%%%%%%%%%%%%%%%%%%
%%% subsection: group theoretic background
%%%%%%%%%%%%%%%%%%%%%%%%%%%%%%%%%%%%%%%%%%%%%%%%%%%%%%%%%%%%%%%%%% 
\subsection{Group theoretic background}
Take an arbitrary $n \in \bN = \{ 1, 2, \ldots \}$ 
and let $N := 2 n$. With the aid of the unitary 
matrix
\be
C := \left( 
\begin{array}{c c}
0 & \1_n \\
\1_n & 0
\end{array}
\right)
\in U(N)
\label{C}
\ee
we define the non-compact real reductive matrix
Lie group
\be
U(n, n) := \{ y \in GL(N, \bC) \, | \, y^* C y = C \}.
\label{U_n_n}
\ee
The fixed-point set of the Cartan involution
$\Theta(y) := (y^{-1})^*$ is the maximal compact 
subgroup
\be
U(n, n)_+ := \{ U \in U(n, n) 
\, | \,
U \mbox{ is unitary} \, \}
\cong U(n) \times U(n),
\ee
meanwhile the submanifold 
$U(n, n)_- 
:= \{ y \in U(n, n) \, | \, \Theta(y) = y^{-1} \}$
consists of the Hermitian elements of $U(n, n)$.
Notice that $C$ (\ref{C}) is central inside 
$U(n, n)_+$.

The Lie algebra of $U(n, n)$ has the form
$\mfu(n, n) 
= \{ X \in \mfgl(N, \bC) \, | \, X^* C + C X = 0 \}$.
The natural trace-pairing 
$\langle X, Y \rangle := \tr(X Y)$ provides an 
invariant, non-degenerate, bilinear form on 
$\mfu(n, n)$. The Lie algebra involution 
$\theta(X) = -X^*$ corresponding to $\Theta$ 
induces the orthogonal $\bZ_2$-gradation
\be
\mfu(n, n) = \mfu(n, n)_+ \oplus \mfu(n, n)_-
\label{gradation}
\ee
with the eigenspaces 
$\mfu(n, n)_\pm := \ker(\theta \mp \Id)$. 
The bilinear form $\langle \, , \rangle$ is negative 
definite on the subalgebra $\mfu(n, n)_+$ and positive 
definite on the complementary subspace $\mfu(n, n)_-$. 
Note that $\mfu(n, n)_+$ (resp. $\mfu(n, n)_-$) consists 
of the anti-Hermitian (resp. Hermitian) elements of 
$\mfu(n, n)$. 

Now with any real $n$-tuple 
$q = (q_1, \ldots, q_n) \in \bR^n$ we associate
the diagonal matrices
\be
\bsq := \diag(q_1, \ldots, q_n) \in \mfgl(n, \bR)
\quad \mbox{and} \quad
Q := \diag(\bsq, -\bsq) \in \mfgl(N, \bR).
\label{q&Q}
\ee
The subset 
$\mfa 
:= \{ Q = \diag(\bsq, -\bsq) \, | \, q \in \bR^n \}$ 
is a maximal Abelian subspace in $\mfu(n, n)_-$. 
Its centralizer inside $U(n, n)_+$ is the Abelian 
group
\be
M := \{ \diag( e^{\ri \bschi}, e^{\ri \bschi} )
\, | \,
\chi \in \bR^n \} \leq U(n, n)_+
\label{M}
\ee
with Lie algebra 
$\mfm := \{ \diag(\ri \bschi, \ri \bschi) 
\, | \, \chi \in \bR^n \} \leq \mfu(n, n)_+$.
Notice that both $\mfa$ and $\mfm$ are realized 
by diagonal matrices. Let $\mfa^\perp$ (resp. 
$\mfm^\perp$) denote the subspace of the 
off-diagonal elements of $\mfu(n, n)_-$ 
(resp. $\mfu(n, n)_+$), then we can write
$\mfu(n, n)_- = \mfa \oplus \mfa^\perp$ and
$\mfu(n, n)_+ = \mfm \oplus \mfm^\perp$.

The subspace $\mfm^\perp \oplus \mfa^\perp$ 
formed by the off-diagonal elements of 
$\mfu(n, n)$ is invariant under the linear 
operator $\ad_Q$, for any $Q \in \mfa$.
Therefore the restricted operator 
\be
\wad_Q := \ad_Q |_{\mfm^\perp \oplus \mfa^\perp}
\label{wad_Q}
\ee
is well-defined, with spectrum 
\be
\sigma(\wad_Q) 
= \{ q_a - q_b, \pm(q_a + q_b), \pm 2 q_c  
\, | \, a, b, c \in \bN_n, a \neq b \},
\ee
where $\bN_n := \{1, \ldots, n \} \subset \bN$.
The regular part of $\mfa$ is defined by 
the subset
\be
\mfa_\reg := \{ Q \in \mfa
\, | \,
\wad_Q \mbox{ is invertible } \}
\subset \mfa.
\label{mfa_reg}
\ee
Since $\mfa \setminus \mfa_\reg$ is a union 
of finitely many hyperplanes, $\mfa_\reg$ 
is open and dense in $\mfa$. The subset 
\be
\mfc := \{ Q = \diag(\bsq, -\bsq)
\, | \,
q = (q_1, \ldots, q_n) \in \bR^n,
q_1 > \ldots > q_n > 0 \}
\label{mfc}
\ee
is a connected component of $\mfa_\reg$, i.e.,
it is an open Weyl chamber. Note that the 
configuration space of the hyperbolic $C_n$ 
Sutherland model (\ref{H}) can be identified with 
$\mfc$. In the following we will frequently use 
the identification 
$\mfc \cong \{ q = (q_1, \ldots, q_n) \in \bR^n 
\, | \, q_1 > \ldots > q_n > 0 \}$. 

As is known, the elements of $\mfu(n, n)_-$ can be 
`diagonalized' by conjugation with elements from 
$U(n, n)_+$. More precisely, let $\mfc^-$ denote the 
closure of $\mfc$, then the map
\be
\mfc^- \times U(n, n)_+ \ni 
(Q, U) \mapsto U Q U^{-1} 
\in \mfu(n, n)_-
\label{u_n_n_-param}
\ee
is well-defined and onto. Moreover,
the regular part of $\mfu(n, n)_-$,
\be
(\mfu(n, n)_-)_\reg := \{ U Q U^{-1}
\, | \,
Q \in \mfc, U \in U(n, n)_+ \} \subset \mfu(n, n)_-,
\ee
is an open and dense subset inside $\mfu(n, n)_-$, 
admitting the smooth bijective parametrization
\be
\mfc \times (U(n, n)_+ / M) \ni 
(Q, U M) \mapsto U Q U^{-1} 
\in (\mfu(n, n)_-)_\reg.
\label{u_n_n_minus_reg_param}
\ee
That is, the above diffeomorphism provides 
the identification 
$(\mfu(n, n)_-)_\reg 
\cong \mfc \times (U(n, n)_+ / M)$.
%%%%%%%%%%%%%%%%%%%%%%%%%%%%%%%%%%%%%%%%%%%%%%%%%%%%%%%%%%%%%%%%
%%% subsection: Lax representation of the Sutherland dynamics
%%%%%%%%%%%%%%%%%%%%%%%%%%%%%%%%%%%%%%%%%%%%%%%%%%%%%%%%%%%%%%%%
\subsection{Lax representation of the Sutherland dynamics}
Let $E \in \bC^N$ denote the column vector with 
components $E_a = 1$, $E_{n + a} = -1$ $(a \in \bN_n)$, 
and set
\be
\xi := \ri g (E E^* - \1_N) + \ri (g - g_2) C 
\in \mfm^\perp \subset \mfu(n, n)_+, 
\label{xi}
\ee
where $g$ and $g_2$ are arbitrary non-zero 
real parameters. Utilizing the Riesz--Dunford 
functional calculus, to any  
$(q, p) \in T^* \mfc \cong \mfc \times \bR^n 
\subset \bR^n \times \bR^n$ we associate the 
$N \times N$ matrix 
\be
\cL(q, p) 
:= P - \coth(\wad_Q) \xi \in \mfu(n, n)_-.
\label{cL}
\ee
Consider the $\mfm$-valued function
$\varPhi := \ri \, \diag(\varphi_1, \ldots, \varphi_n, 
\varphi_1, \ldots, \varphi_n)$
with 
\be
\varphi_c(q, p) =
- g \sum_{a \in \bN_n \setminus \{ c \}} 
\left(
\sinh(q_c - q_a)^{-2} 
+ \sinh(q_c + q_a)^{-2}
\right)
- g_2 \sinh(2 q_c)^{-2},
\ee
and define the $N \times N$ matrix
\be
\cB(q, p) 
:= \varPhi(q, p) 
+ \sinh(\wad_{Q})^{-2} \xi \in \mfu(n, n)_+.
\label{cB}
\ee
As is known, the matrix-valued functions $\cL$ 
and $\cB$ provide a Lax pair for the hyperbolic 
$C_n$ Sutherland model. More precisely, along a  
smooth regular curve $q(t) \in \mfc$ $(t \in \bR)$ 
the Lax equation $\dot{\cL} = [ \cL, \cB ]$
is satisfied if and only if $q(t)$ is a solution 
of the hyperbolic $C_n$ Sutherland dynamics. 

An obvious consequence of the Lax representation 
of the dynamics is that the solution curves can  
be realized as projections of certain geodesics 
on the Riemannian manifold $U(n, n)_-$. Indeed, 
take an arbitrary solution $q(t) \in \mfc$ of the 
Sutherland dynamics and set $p(t) := \dot{q}(t)$. 
The differential equation
$u(t)^{-1} \dot{u}(t) = \cB(q(t), p(t))$
has a unique smooth solution $u(t) \in U(n, n)_+$
$(t \in \bR)$ with initial condition, say, 
$u(0) = \1_N$. Now for any $t \in \bR$ we define 
the positive definite matrix
\be
y(t) := u(t) e^{2 Q(t)} u(t)^{-1} \in U(n, n)_-.
\label{y_def}
\ee 
Clearly $y(t)$ is a smooth function of $t$ 
satisfying the equation
$y^{-1} \dot{y} + \dot{y} y^{-1} = 4 u \cL u^{-1}$.
It follows that $y(t)$ is a solution of the 
\emph{geodesic equation} on $U(n, n)_-$, i.e.,
\be
\frac{\mathrm{d}}{\mathrm{d} t}
\left(
\frac{y^{-1} \dot{y} + \dot{y} y^{-1}}{4}
\right)
= u ( \dot{\cL} - [ \cL, \cB ] ) u^{-1} = 0, 
\label{geodesic_equation}
\ee
with $y(0) = e^{2 Q(0)}$ and 
$y(0)^{-1} \dot{y}(0) + \dot{y}(0) y(0)^{-1} 
= 4 \cL(q(0), p(0))$.
However, upon introducing 
\be
L(q, p) := \cosh(\ad_{Q})^{-1} \cL(q, p)
= P - \sinh(\wad_{Q})^{-1} \xi \in \mfu(n, n)_-,
\label{L}
\ee
the \emph{unique} solution of the geodesic equation 
with the above initial conditions is the curve
\be
y(t) 
= e^{Q(0)} e^{2 t L(q(0), p(0))} e^{Q(0)} .
\label{matrix_flow}
\ee
Comparing (\ref{y_def}) and (\ref{matrix_flow}) 
we see that $Q(t)$, and so the trajectory $q(t)$, 
can be recovered by diagonalizing the matrix flow
(\ref{matrix_flow}). In particular, we have the 
spectral identification
\be
\{e^{2 q_1(t)}, \ldots, e^{2 q_n(t)},
e^{-2 q_n(t)}, \ldots, e^{-2 q_1(t)}\}
= \sigma( e^{2 Q(0)} e^{2 t L(q(0), p(0))} ),
\label{spectral_identification}
\ee
whence the temporal asymptotics of the trajectory
$q(t)$ can be understood by analyzing the 
temporal asymptotics of the eigenvalues of 
the matrix flow (\ref{matrix_flow}). 
Though it is a natural matrix analytic question, 
to our knowledge the first reference containing 
the solution of this problem is 
Ruijsenaars' paper \cite{RuijCMP1988}. 
%%%%%%%%%%%%%%%%%%%%%%%%%%%%%%%%%%%%%%%%%%%%%%%%%%%%
%%% SECTION: TEMPORAL ASYMPTOTICS
%%%%%%%%%%%%%%%%%%%%%%%%%%%%%%%%%%%%%%%%%%%%%%%%%%%%
\section{Temporal asymptotics}
In this section we work out the temporal 
asymptotics of the hyperbolic $C_n$ Sutherland 
dynamics. Our main guide is 
Ruijsenaars' result on the temporal asymptotics 
of the eigenvalues of exponential matrix flows 
(\ref{matrix_flow}). As dictated by Theorem A2 
in \cite{RuijCMP1988}, the plan is to find 
the matrix entries of $e^{2 Q(0)}$ in an 
orthonormal basis, in which $L(q(0), p(0))$ 
is diagonal with decreasing diagonal entries. 
Having control over the matrix entries of 
$e^{2 Q(0)}$ in this new basis, simply by 
computing the quotients of the consecutive 
leading principal minors, we can determine 
the temporal asymptotics of the eigenvalues 
of the matrix flow. 

In the rest of the paper we simply write $L$ and 
$Q$ in place of $L(q(0), p(0))$ and $Q(0)$. 
Since the initial conditions $q(0)$ and 
$p(0) = \dot{q}(0)$ can be arbitrary, we think of 
$L$ and $Q$ as matrix-valued smooth functions 
over the phase space $T^*\mfc \cong \mfc \times \bR^n$. 
%%%%%%%%%%%%%%%%%%%%%%%%%%%%%%%%%%%%%%%%%%%%%%%%%%%%%%%%%
%%% subsection: Diagonalization of L  
%%%%%%%%%%%%%%%%%%%%%%%%%%%%%%%%%%%%%%%%%%%%%%%%%%%%%%%%%  
\subsection{Analyzing the spectrum of $L$}
Recall that it is a crucial assumption of 
Theorem A2 in \cite{RuijCMP1988} that the
spectrum of matrix $L$ is simple. To examine
the spectrum of $L$, we set up an equation 
for $L \in \mfu(n, n)_-$ and 
$A := e^{2 Q} \in U(n, n)_-$ as follows. 
By applying the linear operator $\sinh(\ad_Q)$ 
on $L$ (\ref{L}), we get 
\be
\sinh(\ad_Q) L = - \xi,
\ee
which entails
$L e^{2 Q} - e^{2 Q} L = 2 e^{Q} \xi e^{Q}$.
Now, recalling (\ref{xi}), we can write
\be
2 \ri g A + L A - A L 
= 2 \ri g (e^Q E) (e^Q E)^*  
+ 2 \ri (g - g_2) C.
\label{L&A_eqn}
\ee
Note that this equation is the complete analogue 
of Ruijsenaars' commutation relation 
(equation (2.4) in \cite{RuijCMP1988})  
he analyzed to discover the remarkable 
\emph{action-angle duality} 
between the hyperbolic $A_n$ Sutherland and the 
rational $A_n$ Ruijsenaars--Schneider models. 
If $g_2 = g$, the right hand side of (\ref{L&A_eqn})
is a matrix of rank one, therefore the analysis of
the equation is relatively straightforward.
Notice that the special case $g_2 = g$ corresponds 
to the $\bZ_2$-folding of the $A_{2 n - 1}$ model. 
However, when $g_2 \neq g$, the innocent looking 
term $2 \ri (g - g_2) C$ complicates the analysis 
considerably.

To proceed further, we diagonalize $L \in \mfu(n, n)_-$. 
As we saw in (\ref{u_n_n_-param}), we can write 
\be
L = U \check{L} U^{-1} = U \check{L} U^*
\label{checkL&U} 
\ee
with some $\check{L} \in \mfc^-$ and $U \in U(n, n)_+$.
Observe that $\check{L}$ is unique, having the form 
$\check{L} = \diag(\bslambda, -\bslambda)$ with some 
$\lambda_1 \geq \ldots \geq \lambda_n \geq 0$,
but the choice of $U$ is not unique. However, at 
this point all we need is the existence of the pair 
$(\check{L}, U)$, the consequences of the 
non-uniqueness will be discussed at the end of this 
subsection. Now from (\ref{L&A_eqn}) we conclude 
that $\check{L}$ and 
$\check{A} := U^{-1} A U \in U(n, n)_-$ satisfy 
the equation
\be
2 \ri g \check{A} + \check{L} \check{A} 
- \check{A} \check{L} 
= 2 \ri g (U^* e^Q E) (U^* e^Q E)^* 
+ 2 \ri (g - g_2) C.
\label{L&A_check_eqn}
\ee
Computationwise it is very fortunate that the matrix 
$\check{A}^{-1} \in U(n, n)_-$ obeys a similar equation. 
Indeed, by conjugating the above equation with $C$, 
we get
\be
2 \ri g \check{A}^{-1} - \check{L} \check{A}^{-1} 
+ \check{A}^{-1} \check{L} 
= 2 \ri g (C U^* e^Q E) (C U^* e^Q E)^* 
+ 2 \ri (g - g_2) C.
\label{L&A_check_inverse_eqn}
\ee
Upon introducing the purely imaginary numbers
\be
x_c = - x_{n + c} := (2 \ri g)^{-1} \lambda_c
\in \ri \bR 
\quad 
(c \in \bN_n),
\ee
the column vector 
\be
F := U^* e^Q E \in \bC^N, 
\label{F}
\ee
and the real parameter 
\be
\varepsilon := 1 - g_2 g^{-1} \in \bR,
\label{varepsilon}
\ee
for the matrix entries of $\check{A}$ and
$\check{A}^{-1}$ we obtain 
\be
\check{A}_{k, l} 
= \frac{F_k \overline{F}_l 
+ \varepsilon C_{k, l}}{1 + x_k - x_l},
\quad
(\check{A}^{-1})_{k, l} 
= \frac{(C F)_k \overline{(C F)}_l 
+ \varepsilon C_{k, l}}{1 - x_k + x_l},
\label{check_A_entries}
\ee
for any $k, l \in \bN_N$. Note that the relations
\be
\sum_{j = 1}^{N}
\frac{F_k \overline{F}_j + \eps C_{k, j}}
{1 + x_k - x_j}
\frac{(C F)_j \overline{(C F)}_l 
+ \eps C_{j, l}}{1 - x_j + x_l}
= \delta_{k, l}
\label{quad_eqn}
\ee 
obviously follow from (\ref{check_A_entries}).

Having equipped with the above formulae, 
we are able to analyze the spectrum of $L$ 
and the properties of column vector $F$.
For convenience, we introduce the notations
\be
f_c := F_c,
\quad
h_c := F_{n + c}, 
\quad
z_c := f_c \overline{h}_c
\quad (c \in \bN_n).
\label{f&h&z}
\ee

\smallskip
\noindent
\textbf{Lemma 1.}
\emph{
If the non-zero coupling parameters $g$ and $g_2$ 
satisfy $g_2 \neq 2 g$, then the components of the 
column vector $F$ are non-zero and $L$ is a regular 
element of $\mfu(n, n)_-$.
}

\medskip
\noindent
\textbf{Proof.}
We only show that the components of $F$ are 
non-zero. Proving by contraposition, suppose 
that $z_c = 0$ for some $c \in \bN_n$. With 
$k = l = c$, from (\ref{quad_eqn}) we get the 
quadratic relation $\eps^2 = (1 + 2 x_c)^2$, 
i.e., $\eps = \pm (1 + 2 x_c)$. By comparing 
the real parts we obtain $\eps = \pm 1$, 
which contradicts our assumption on the 
coupling parameters. Along the same line, 
by appropriately specializing the indices in 
equation (\ref{quad_eqn}), the regularity 
of $L$ also follows. 
\hspace*{\stretch{1}} \qedsymb

\noindent
\textbf{Remark.}
Henceforth we assume that the parameters $g$ 
and $g_2$ satisfy the additional technical 
condition $g_2 \neq 2 g$. Notice, however, 
that it \emph{does not} restrict the values 
of the physically relevant positive coupling 
constants $g^2$ and $g_2^2$, since the pairs
$(g, g_2)$ and $(g, -g_2)$ generate the same 
couplings in the model (\ref{H}). In principle, 
without loss of generality, we could have imposed 
the condition $g g_2 < 0$ at the outset, thereby 
automatically excluding the case $g_2 = 2 g$. 

To conclude this subsection we wish to point 
out that the construction of $z_c$ (\ref{f&h&z}) 
results in a well-defined smooth function on the
phase space $T^* \mfc$. As we saw in 
(\ref{u_n_n_minus_reg_param}), by the regularity of 
$L$, the non-uniqueness of the diagonalizing matrix 
$U \in U(n, n)_+$ defined in (\ref{checkL&U}) is 
controlled entirely by the centralizer subgroup 
$M$ (\ref{M}). Namely, the only freedom in the 
choice of $U$ can be characterized by the 
transformations
\be
U \mapsto U \diag(e^{\ri \bschi}, e^{\ri \bschi}),
\ee
generated by some $\chi \in \bR^n$. Now observe that
the components $f_c$ and $h_c$ (\ref{f&h&z}) 
of the column vector $F$ (\ref{F}) transform as
\be
f_c \mapsto e^{-\ri \chi_c} f_c 
\quad \mbox{and} \quad 
h_c \mapsto e^{-\ri \chi_c} h_c, 
\label{f&h_transformation}
\ee
hence $z_c = f_c \overline{h}_c$ is independent 
of the choice of the representative $U$.
It means that to each $L = L(q, p)$ we can 
associate the non-zero complex numbers 
$z_c = z_c(q, p)$ $(c \in \bN_n)$ in a unique 
and well-defined manner. To show that their 
dependence on the phase space variables 
$(q, p)$ is smooth, we notice that by choosing 
appropriate \emph{smooth local sections} of 
the (smooth) fiber bundle 
\be
\mfc \times U(n, n)_+ 
\twoheadrightarrow 
\mfc \times (U(n, n)_+ / M)
\cong (\mfu(n, n)_-)_\reg, 
\ee
we can work with representatives $U \in U(n, n)_+$ 
depending smoothly on the phase space variables in 
a small neighborhood of any given 
$(q, p) \in T^* \mfc$. Thus, $z_c$ is smooth
around any $(q, p)$, proving its smoothness over 
the whole phase space.
%%%%%%%%%%%%%%%%%%%%%%%%%%%%%%%%%%%%%%%%%%%%%%%%%%%%%%%%%
%%% subsection: On the structure of check A
%%%%%%%%%%%%%%%%%%%%%%%%%%%%%%%%%%%%%%%%%%%%%%%%%%%%%%%%%
\subsection{The structure of $\check{A}$}
In this subsection we proceed with a detailed 
analysis on the structure of matrix $\check{A}$. 
To this end, we make use of Jacobi's theorem in linear 
algebra, i.e., we exploit some non-trivial relations 
between certain minors of matrices  $\check{A}$ and 
$\check{B} := (\check{A}^{-1})^T$.
For convenience, in Appendix A we provide a brief 
account on the relevant theorems from linear algebra.

In the following we keep the index $c \in \bN_n$ 
fixed and apply Jacobi's theorem on appropriate 
minors of $\check{A}$ and $\check{B}$. 
Since $\det(A) = \det(e^{2 Q}) = 1$, let us keep
in mind that $\det(\check{A}) = 1$ also holds. 
To make the presentation shorter we introduce the 
notations
\be
\cD_c :=  
\prod_{\substack{ d = 1 \\ (d \neq c)}}^n 
\vert h_d \vert^2 
\prod_{\substack{a, b = 1 \\ (c \neq a \neq b \neq c)}}^n
\frac{x_a - x_b}{1 + x_a - x_b} 
\quad \mbox{and} \quad
\omega_c := \prod_{\substack{a = 1 \\ (a \neq c)}}^n 
\frac{(x_c - x_a)(x_c + x_a)}
{(1 + x_c - x_a)(1 + x_c + x_a)}.
\label{D&omega}
\ee
Clearly $\cD_c \in \bR \setminus \{ 0 \}$   
and $\omega_c \in \bC \setminus \{ 0 \}$.
Note also that in the equations below the 
matrices $e_{k, l}$ stand for the elementary 
matrices, i.e., their entries are defined as 
$(e_{k, l})_{k', l'} = \delta_{k, k'} \delta_{l, l'}$.

As a first application of Jacobi's theorem, 
we can write
\be
\check{B} \left(
\begin{array}{c c c c c}
1 & \cdots & c & \cdots & n \\
1 & \cdots & n + c & \cdots & n
\end{array} \right)
= - \check{A} \left(
\begin{array}{c c c c c c}
n + 1 & \cdots & n + c & \cdots & 2 n \\
n + 1 & \cdots & c & \cdots & 2 n
\end{array} \right).
\label{Jac_1}
\ee 
Let $R$ and $S$ denote the $n \times n$ submatrices
corresponding to the above minors on the left and on 
the right, respectively. Upon introducing the 
Cauchy-type $n \times n$ matrix $\Psi$ with entries
\be
\Psi_{a, b} 
:= \frac{\overline{h}_a h_b}{1 + x_a - x_b}  
\mbox{ if $b \neq c$, and } 
\Psi_{a, c} 
:= \frac{\overline{h}_a f_c}{1 + x_a - x_{n + c}}, 
\label{Psi}
\ee
from (\ref{check_A_entries}) we see that 
\be
R = \Psi + \eps (1 + 2 x_c)^{-1} e_{c, c},
\ee
meanwhile the entries of $S$ can be identified as
$S_{a, b} = \overline{R}_{a, b}$ $(a, b \in \bN_n)$. 
Therefore, equation (\ref{Jac_1}) can be cast into 
the particularly simple form 
\be
\det(R) + \overline{\det(R)} = 0.
\label{det_eqn}
\ee
Since $R$ is a rank one perturbation of 
$\Psi$, the determinant formula 
(\ref{rank_1_formula}) spells out as 
\be
\det(R) = \det(\Psi) + \eps (1 + 2 x_c)^{-1} \cC_{c, c},
\ee
where $\cC_{c, c}$ is the cofactor of $\Psi$ 
associated with entry $\Psi_{c, c}$. Since 
$\Psi$ is of Cauchy-type, we get
\be 
\det(\Psi) = (1 + 2 x_c)^{-1} \cD_c \omega_c z_c
\quad \mbox{and} \quad
\cC_{c, c} = \cD_c.
\ee 
Plugging these formulae into (\ref{det_eqn}), 
for $z_c$ we obtain the linear relation
\be
(1 - 2 x_c) \omega_c z_c 
+ (1 + 2 x_c) \overline{\omega}_c \overline{z}_c 
+ 2 \varepsilon = 0.
\label{z_c_linear_eqn}
\ee
Notice that this single equation does not
determine uniquely the complex quantity $z_c$.

To get an independent relation for $z_c$, we
turn to Jacobi's theorem, again. Namely, we 
can write 
\be
\check{B} \left(
\begin{array}{c c c c}
1 & \cdots & n & n + c \\
1 & \cdots & n & n + c
\end{array} \right)
= \check{A} \left(
\begin{array}{c c c c c}
n + 1 & \cdots & \widehat{n + c} & \cdots & 2 n \\
n + 1 & \cdots & \widehat{n + c} & \cdots & 2 n
\end{array} \right),
\label{Jac_2}
\ee 
where the symbol $\widehat{n + c}$ means that the
indicated row (and column) is \emph{deleted} 
in the minor on the right hand side. 
That is, on the left we have a principal minor 
of $\check{B}$ of size $n + 1$, and the 
principal minor of $\check{A}$ on the right has 
size $n - 1$. Let $X$ and $Y$ denote the submatrices 
corresponding to these minors, respectively, 
then we have $\det(X) = \det(Y)$. 
Since $Y$ is of Cauchy-type, the relation
$\det(Y) = \cD_c$ immediately follows. On the other 
hand, the computation of $\det(X)$ requires a 
longer preparation. To this end, we introduce the 
Cauchy-type $(n + 1) \times (n + 1)$ matrix 
$\Phi$ with entries
\be
\Phi_{a, b} 
:= \frac{\overline{h}_a h_b}{1 + x_a - x_b},
\;
\Phi_{a, n + 1} 
:= \frac{\overline{h}_a f_c}{1 + x_a - x_{n + c}},
\;
\Phi_{n + 1, b} 
:= \frac{\overline{f}_c h_b}{1 + x_{n + c} - x_b},
\;
\Phi_{n + 1, n + 1} 
:= \vert f_c \vert^2, 
\ee
where $a, b \in \bN_n$. Recalling 
(\ref{check_A_entries}), we see that
\be
X = \Phi 
+ \eps (1 + 2 x_c)^{-1} e_{c, n + 1}  
+ \eps (1 - 2 x_c)^{-1} e_{n + 1, c},
\label{X}
\ee
i.e., $X$ is a rank two perturbation 
of $\Phi$. Therefore, the determinant 
formula (\ref{rank_2_formula}) yields
\be
\det(X) = \det(\Phi)
+ \eps  \left(
\frac{\cC_{c, n + 1}}{1 + 2 x_c} 
+ \frac{\overline{\cC}_{c, n + 1}}{1 - 2 x_c} 
\right) 
+ \eps^2 
\frac{\vert \cC_{c, n + 1} \vert^2 
- \cC_{c, c} \cC_{n + 1, n + 1}}
{(1 - 4 x_c^2)\det(\Phi)}, 
\ee
where the $\cC_{k, l}$'s now denote the cofactors 
of $\Phi$. Using the special Cauchy-type form 
of $\Phi$, we obtain
\be
\det(\Phi) 
= -4 x_c^2 (1 - 4 x_c^2)^{-1} 
\cD_c \vert \omega_c z_c \vert^2,
\quad
\cC_{c, n + 1} = -(1 - 2 x_c)^{-1} 
\cD_c \overline{\omega}_c \overline{z}_c, 
\ee
together with the relations
\bea
&& \cC_{c, c} = \cD_c \vert f_c \vert^2 
\prod_{\substack{a = 1 \\ (a \neq c)}}^n 
\frac{(x_c + x_a)(-x_c - x_a)}
{(1 + x_c + x_a)(1 - x_c - x_a)}, \\
&& \cC_{n + 1, n + 1} = \cD_c \vert h_c \vert^2 
\prod_{\substack{a = 1 \\ (a \neq c)}}^n 
\frac{(x_c - x_a)(-x_c + x_a)}
{(1 + x_c - x_a)(1 - x_c + x_a)}.
\eea
It immediately follows that the determinant 
of $X$ has the form 
\be
\det(X) = - (1 - 4 x_c^2)^{-1} \cD_c
\left( 4 x_c^2 \vert \omega_c z_c \vert^2 
+ \varepsilon (\omega_c z_c 
+ \overline{\omega}_c \overline{z}_c)
+ \varepsilon^2 \right).
\label{det_X}
\ee
Finally, by putting these formulae together, 
we end up with the quadratic equation
\be
4 x_c^2 \vert \omega_c z_c \vert^2 
+ \eps (\omega_c z_c 
+ \overline{\omega}_c \overline{z}_c)
+ \eps^2  + 1 - 4 x_c^2 = 0.
\label{z_c_quadratic_eqn}
\ee

Next, by solving equations 
(\ref{z_c_linear_eqn}) and 
(\ref{z_c_quadratic_eqn})
for $z_c$, we find the following two 
solutions 
\be
z_c 
= \pm \left( 1 + \frac{1 \pm \eps}{2 x_c} \right) 
\prod_{\substack{a = 1 \\ (a \neq c)}}^n
\left(1 + \frac{1}{x_c - x_a} \right)
\left(1 + \frac{1}{x_c + x_a} \right).
\label{z_c_non_unique}
\ee
In order to select the right one, we proceed as
follows. In the phase space region where the 
particles are far from each other, i.e.,
$q_1 \gg \ldots \gg q_n \gg 0$,
moving with high relative momenta, i.e.,
$p_1 \gg \ldots \gg p_n \gg 0$,
the matrix $L$ (\ref{L}) is almost diagonal, 
therefore the diagonalizing matrix $U$
(\ref{checkL&U}) is also nearly diagonal. 
Recalling equations (\ref{F}) and (\ref{f&h&z}), 
we see that in the given phase space region 
the value of $z_c$ is very close to $-1$. 
Since $z_c$ is a \emph{smooth function}  
over the \emph{connected} phase space 
$T^* \mfc$, by invoking a standard continuity 
argument, we conclude that 
\be
z_c = - \left( 1 + \frac{\ri g_2}{\lambda_c} \right) 
\prod_{\substack{a = 1 \\ (a \neq c)}}^n
\left(1 + \frac{2 \ri g}{\lambda_c - \lambda_a} \right)
\left(1 + \frac{2 \ri g}{\lambda_c + \lambda_a} \right).
\label{z_c}
\ee

Having determined $z_c$, we can find the 
form of the components $f_c$ and $h_c$ 
(\ref{f&h&z}), too. As we discussed at the 
end of the previous subsection, 
by the non-uniqueness of $U$ (\ref{checkL&U}),
the non-zero complex quantities $f_c$ and 
$h_c$ are determined only up to a common phase 
factor (\ref{f&h_transformation}). Therefore, 
purely for convenience, we may and shall assume 
that $f_c > 0$ for any $c \in \bN_n$. So, we can 
write
\be 
f_c = e^{\theta_c} \vert z_c \vert^{\frac{1}{2}}
\quad \mbox{and} \quad
h_c = e^{-\theta_c} \overline{z}_c 
\vert z_c \vert^{-\frac{1}{2}}
\label{f&h_parametrization}
\ee
with some $\theta_c \in \bR$. 
Combining this parametrization with 
(\ref{check_A_entries}), we obtain the 
following description of the matrix 
$\check{A}$.

\smallskip
\noindent
\textbf{Lemma 2.}
\emph{
With the aid of the $\lambda$-dependent 
functions $z_c$ (\ref{z_c}), the matrix 
entries of $\check{A}$ take the form
\bea
&& \check{A}_{a, b} 
= e^{\theta_a + \theta_b} 
\vert z_a z_b \vert^\frac{1}{2}
\frac{2 \ri g}{2 \ri g + \lambda_a - \lambda_b}, 
\quad
\check{A}_{n + a, n + b} 
= e^{-\theta_a - \theta_b} 
\frac{\overline{z}_a z_b}
{\vert z_a z_b \vert^{\frac{1}{2}}}
\frac{2 \ri g}{2 \ri g - \lambda_a + \lambda_b}, 
\label{Lemma_2_1} \\
&& \check{A}_{a, n + b} = \overline{\check{A}}_{n + b, a} 
= e^{\theta_a - \theta_b} z_b 
\vert z_a z_b^{-1} \vert^\frac{1}{2}
\frac{2 \ri g}{2 \ri g + \lambda_a + \lambda_b}
+ \frac{\ri(g - g_2)}{\ri g + \lambda_a} \delta_{a, b},
\label{Lemma_2_2}
\eea
where $a, b \in \bN_n$. 
}

\medskip
\noindent
\textbf{Remark.}
Due to the presence of the $g_2$-dependent 
second term on the right-hand side of equation 
(\ref{Lemma_2_2}), the matrix $\check{A}$ can 
be seen as a $C_n$-type non-trivial 
\emph{deformation} of the usual Cauchy matrices. 
Spelling out the relation $\det(\check{A}) = 1$, 
the resulting Cauchy-type determinant formula 
might be of interest in other branches 
of mathematics and physics as well.
%%%%%%%%%%%%%%%%%%%%%%%%%%%%%%%%%%%%%%%%%%%%%%%%%%%%%%
%%% subsection: Asymptotic phases and momenta
%%%%%%%%%%%%%%%%%%%%%%%%%%%%%%%%%%%%%%%%%%%%%%%%%%%%%%
\subsection{Asymptotic phases and momenta}
Now let $q(t) = (q_1(t), \ldots, q_n(t)) \in \mfc$ 
be an arbitrary solution of the hyperbolic $C_n$
Sutherland dynamics, then by 
(\ref{spectral_identification}) we can write
\be
\{ e^{2 q_1(t)}, \ldots, e^{2 q_n(t)}, 
e^{-2 q_n(t)}, \ldots, e^{-2 q_1(t)} \} 
= \sigma(\check{A} e^{2 t \check{L}}),
\label{spectral_identification_2}
\ee
where both $L$ and $A$ are computed at time 
$t = 0$. Since the diagonal entries of 
$\check{L} = \diag(\bslambda, -\bslambda)$ 
are not in decreasing order, we conjugate 
it by the unitary $N \times N$ matrix 
$W := \diag(\1_n, \cR_n)$, where $\cR_n$ 
is the unitary $n \times n$ matrix with entries 
$(\cR_n)_{a, b} 
:= \delta_{a + b, n + 1}$ $(a, b \in \bN_n)$.
Upon setting 
\be
\hat{L} := W \check{L} W^{-1}
\quad \mbox{and} \quad
\hat{A} := W \check{A} W^{-1}, 
\ee
we see that the diagonal entries of
$\hat{L} = \diag(\lambda_1, \ldots, \lambda_n, 
-\lambda_n, \ldots, -\lambda_1)$
are decreasing, and for any $a, b \in \bN_n$
we have
\be
\hat{A}_{a, b} = \check{A}_{a, b} 
= f_a (1 + x_a - x_b)^{-1} \overline{f}_b.
\ee
At this point Ruijsenaars' theorem 
\cite{RuijCMP1988} on the   
temporal asymptotics of exponential matrix 
flows is directly applicable. Recalling 
(\ref{spectral_identification_2}), for 
any $c \in \bN_n$ we obtain the asymptotic 
relation
\be
e^{2 q_c(t)} \sim m_c e^{2 t \lambda_c} 
= e^{ \ln(m_c) + 2 t \lambda_c}
\quad
(t \to \infty),
\ee
where the $m_c$'s stand for the quotients      
of the consecutive leading principal minors 
of $\hat{A}$. Since the $c$th leading principal 
minor of $\hat{A}$ has the form
\be
\cP_c := \prod_{d = 1}^c \vert f_d \vert^2 
\prod_{1 \leq a < b \leq c} 
\left(
1 - (x_a - x_b)^{-2}
\right)^{-1},
\ee
we get
\be
m_c = \frac{\cP_c}{\cP_{c - 1}} 
= \vert f_c \vert^2 \prod_{a = 1}^{c - 1}
\left(
1 - (x_a - x_c)^{-2}
\right)^{-1}. 
\ee
Therefore, for $t \to \infty$ we have 
$q_c(t) \sim q_c^+ + t p_c^+$ with asymptotic
phases and momenta
\be
q_c^+ = \frac{1}{2} \ln(m_c) 
= \frac{1}{2} \ln(\vert f_c \vert^2) 
- \frac{1}{2} \sum_{a = 1}^{c - 1} \ln
\left(
1 - (x_c - x_a)^{-2} 
\right)
\quad \mbox{and} \quad
p_c^+ = \lambda_c. 
\label{t+}
\ee

By conjugating both $\hat{L}$ and $\hat{A}$ 
with the unitary $N \times N$ matrix $\cR_N$,
we see that the diagonal entries of 
$\cR_N \hat{L} \cR_N^{-1}$ are in strictly 
increasing order, so the $t \to -\infty$ 
asymptotics can be handled similarly to the 
$t \to \infty$ case. It turns out that for 
$t \to -\infty$ we have 
$q_c(t) \sim q_c^- + t p_c^-$, where 
\be
q_c^- = \frac{1}{2} \ln(\vert h_c \vert^2) 
- \frac{1}{2} \sum_{a = 1}^{c - 1} \ln
\left(
1 - (x_c - x_a)^{-2}
\right)
\quad \mbox{and} \quad
p_c^- = -\lambda_c. 
\label{t-}
\ee

Now we are in a position to formulate the
main result of the paper. Indeed, recalling
the form of $z_c$ (\ref{z_c}), the comparison 
of (\ref{t+}) and (\ref{t-}) immediately 
leads to the precise relationships between 
the asymptotic phases and momenta.

\smallskip
\noindent
\textbf{Theorem 3.}        
\emph{
Take an arbitrary solution
$q(t) \in \mfc$ $(t \in \bR)$
of the hyperbolic $C_n$ Sutherland dynamics.
For $\vert t \vert \to \infty$ the
particles move asymptotically freely, i.e.,
for any $c \in \bN_n$ we have the asymptotics
\be
q_c(t) \sim q_c^\pm + t p_c^\pm
\quad
(t \to \pm \infty).
\ee 
The asymptotic momenta satisfy the relations 
\be
p_c^+ = -p_c^- 
\quad \mbox{and} \quad
p_1^+ > \ldots > p_n^+ > 0, 
\ee
and for the asymptotic phases we have 
\be
q_c^+ 
= -q_c^-
- \sum_{a = 1}^{c - 1} \delta(p_c^{-} - p_a^{-}, g)
+ \sum_{a = c + 1}^{n} \delta(p_c^{-} - p_a^{-}, g)
+ \sum_{\substack{a = 1 \\ (a \neq c)}}^{n} 
\delta(p_c^{-} +p_a^{-}, g)
+ \delta(2 p_c^{-}, g_2)
\ee
with the $2$-particle phase shift function 
$\delta(p, \mu) = 2^{-1} \ln(1 + 4 \mu^2 p^{-2})$.
} 
%%%%%%%%%%%%%%%%%%%%%%%%%%%%%%%%%%%%%%%%%%%%%%%%%%%%%%
%%% section: Discussion
%%%%%%%%%%%%%%%%%%%%%%%%%%%%%%%%%%%%%%%%%%%%%%%%%%%%%%
\section{Discussion}
In this paper we examined the scattering 
properties of the hyperbolic $C_n$ Sutherland 
model. Under certain technical conditions
the classical repulsive particle systems are 
Liouville integrable, having only scattering 
states. Indeed, in general the asymptotic 
momenta provide sufficiently many independent
first integrals in involution.
However, the asymptotic momenta of 
the hyperbolic $C_n$ Sutherland model satisfy 
also the peculiar algebraic conditions
$p_c^+ = -p_c^-$ $(c \in \bN_n)$. 
Following Ruijsenaars' terminology 
\cite{RuijFiniteDimSolitonSystems},
this distinguishing feature gives grounds for 
calling the $C_n$-type model a 
\emph{pure soliton system}.
As in the $A_n$-type models, we expect that 
this stronger notion of integrability is 
responsible for the factorized form of the 
scattering map. 

We find it interesting that the application
of elementary linear algebraic techniques 
succeeds in revealing the scattering  
behavior of the $C_n$ model. Since the 
hyperbolic $BC_n$ Sutherland model
with three independent coupling constants 
is also closely tied with the matrix Lie 
group $U(n, n)$ (see \cite{FeherPusztai2007}), 
we believe that the algebraic machinery 
presented in this paper can be extended 
to understand the scattering properties 
of the $BC_n$ model, too. Nevertheless, a 
complete classification of the pure soliton 
systems associated with the $BC_n$ root system 
appears to be a more challenging analytic 
problem. For further motivation we mention 
that these finite dimensional many-particle 
systems are closely connected with integrable
field theories. For example, it is well-known 
that the family of $A_n$-type Ruijsenaars--Schneider 
models describe the soliton solutions of the 
sine-Gordon equation (see e.g. \cite{RuijSchneider}, 
\cite{BabelonBernard}). 
Following \cite{KapustinSkorik}, it is also natural 
to speculate on the mathematically adequate description 
of the relationship between the scattering theory of 
the $BC_n$-type particle systems and the soliton 
dynamics of the boundary sine-Gordon models, at both 
the classical and the quantum level. We 
wish to come back to these issues in later 
publications. 

To explore the scattering properties
of the $C_n$ model we adapted Ruijsenaars'
approach \cite{RuijCMP1988} to the $C_n$
root system. Thus, it is quite natural 
that the matrix $\check{A}$ appearing in 
Lemma 2 can be interpreted as the Lax matrix 
of the rational $C_n$ RSvD model. Recall that 
matrix $\check{A}$ depends on $2 n$ real 
parameters 
$\lambda = (\lambda_1, \ldots, \lambda_n)$
and $\theta = (\theta_1, \ldots, \theta_n)$
with $\lambda_1 > \ldots > \lambda_n > 0$.
Regarding the parameters $\lambda_c$ and 
$\theta_c$ as canonically conjugated positions 
and momenta, the natural Hamiltonian associated 
with the matrix $\check{A}$ takes the form
\bea
\lefteqn{H(\lambda, \theta) 
= \frac{1}{4} \tr( \check{A} + \check{A}^{-1} ) } \nonumber\\
&& = \sum_{c = 1}^{n}
\cosh(2 \theta_c)
\left( 1 
+ \frac{g_2^2}
{\lambda_c^2} \right)^\frac{1}{2}
\prod_{\substack{a = 1 \\ (a \neq c)}}^{n}
\left( 1 
+ \frac{4 g^2}
{(\lambda_c - \lambda_a)^2} \right)^\frac{1}{2}
\left( 1 
+ \frac{4 g^2}
{(\lambda_c + \lambda_a)^2} \right)^\frac{1}{2}.
\label{H_RSvD}
\eea
Now let us observe that, up to an irrelevant additive 
constant, this function can be identified with 
the rational limit of van Diejen's Hamiltonian 
(see equation (6) in \cite{vanDiejen1994}) with 
two independent coupling parameters. To prove 
the canonicity of $\lambda_c$ and $\theta_c$ 
and to show the duality between the hyperbolic 
$C_n$ Sutherland and the rational $C_n$ 
RSvD models, one could imitate Ruijsenaars' original 
work \cite{RuijCMP1988} on the systems of type $A_n$. 
However, for the $A_n$-type models it has been shown 
recently \cite{FeherKlimcik0901} that 
the symplectic reduction framework 
provides the most efficient way to establish        
the duality relation. For completeness, we now briefly 
outline the reduction picture underlying the 
duality of the $C_n$ models. The compact Lie 
group $U(n, n)_+$ naturally acts on the manifold 
$U(n, n)_-$ by conjugations. The lift of this 
action to the cotangent bundle $T^* U(n ,n)_-$ 
is Hamiltonian, admitting an equivariant momentum 
map $J \colon T^* U(n, n)_- \rightarrow
\mfu(n, n)_+^* \cong \mfu(n, n)_+$.
In the reduction picture our functional 
equation (\ref{L&A_eqn}) corresponds to 
the momentum map constraint $J = \xi$ 
with the special choice $\xi$ (\ref{xi}). 
Solving this constraint in a gauge
in which $A$ is diagonal, the outcome of
the reduction procedure is the usual phase 
space of the Sutherland model (see e.g.
\cite{OlshaPere}, \cite{AvanBabelonTalon1994},
\cite{FeherPusztai2006}), and the Sutherland
dynamics (\ref{H}) is induced by the Hamiltonian
$H = \tr(L^2) / 4$. Note, however, that in Section 3 
we solved the momentum map constraint
in a gauge in which $L$ is \emph{diagonal}. 
As a result, the reduced phase space can be
parametrized by the $\lambda_c$'s and the 
$\theta_c$'s, and the dynamics of our interest 
(\ref{H_RSvD}) is induced by the Hamiltonian 
$H = \tr(A + A^{-1}) / 4$. 
Therefore, by using two different gauges, 
one in which $A$ is diagonal, and one in which 
$L$ is diagonal, we obtain two different
realizations of the reduced phase space 
$T^* U(n, n)_- //_\xi \, U(n, n)_+$, 
whence a natural symplectomorphism between 
the phase spaces of the hyperbolic $C_n$ 
Sutherland and the rational $C_n$ 
RSvD models comes for free. The details, 
together with the generalization to the 
$BC_n$ systems, will be published elsewhere.

Notice that the action-angle duality between
the Sutherland and the RSvD models has important 
consequences from the perspective of scattering 
theory, too. Based on the duality relation
the scattering theory of the rational 
$BC_n$ RSvD models could also be understood. 
Moreover, as in the $A_n$-type models, the duality 
could greatly simplify the verification of the 
simplecticity of the M\o ller wave transformations 
both for the Sutherland and the RSvD models, 
thereby the understanding of their scattering 
theory would be complete.
%%%%%%%%%%%%%%%%%%%%%%%%%%%%%%%%%%%%%%%%%%%%%%%%%%%%%%
%%% section: Appendix
%%%%%%%%%%%%%%%%%%%%%%%%%%%%%%%%%%%%%%%%%%%%%%%%%%%%%%
\renewcommand{\thesection}{A}
\section{Appendix}
\renewcommand{\theequation}{A.\arabic{equation}}
\setcounter{equation}{0}
In this appendix we summarize some linear 
algebraic facts used throughout the paper. 
The proofs and further details can be found 
e.g. in \cite{Prasolov}. 

For an $n \times n$ matrix $X$, let
\be
X \left(
\begin{array}{c c c c}
k_1 & k_2 & \cdots & k_p \\
l_1 & l_2 & \cdots & l_p 
\end{array} \right)
\ee
denote the minor determinant of the $p \times p$ 
submatrix of $X$ lying on the intersection of 
rows $k_1$, $k_2$, $\ldots$, $k_p$ with columns
$l_1$, $l_2$, $\ldots$, $l_p$.
Jacobi's theorem claims that there are simple 
relationships between the minors of $X$ and the 
minors of its inverse, as described below.

\smallskip
\noindent
\textbf{Theorem A1.}
\emph{
Let $X$ be an invertible $n \times n$ matrix,
$Y := (X^{-1})^T$, and choose a permutation
\be
\sigma = \left(
\begin{array}{c c c c}
k_1 & k_2 & \cdots & k_n \\
l_1 & l_2 & \cdots & l_n 
\end{array} 
\right) \in S_N
\ee
of the pairwise distinct indices 
$k_1, k_2, \ldots, k_n \in \bN_n$.
Then for any $p \in \{0, 1, \ldots, n \}$ 
we have
\be
Y \left(
\begin{array}{c c c c}
k_1 & k_2 & \cdots & k_p \\
l_1 & l_2 & \cdots & l_p 
\end{array} \right)
= \frac{\sgn(\sigma)}{\det(X)} 
X \left(
\begin{array}{c c c c}
k_{p + 1} & k_{p + 2} & \cdots & k_n \\
l_{p + 1} & l_{p + 2} & \cdots & l_n 
\end{array} \right),
\ee
where $\sgn(\sigma)$ denotes the sign of 
permutation $\sigma$.
}

\medskip
\noindent
In this paper we frequently encounter Cauchy
matrices and their perturbations. Recall that
for the determinants of the Cauchy matrices 
we have 
\be
\det \left( 
\frac{1}{1 + \xi_k - \eta_l} 
\right)
= \frac{\prod_{k < l} (\xi_k - \xi_l) (\eta_l - \eta_k)}
{\prod_{k, l} (1 + \xi_k - \eta_l)},
\ee
where the $\xi_k$'s and the $\eta_l$'s are arbitrary 
complex numbers. The key formula that allows us to
compute effectively the determinants of perturbed 
matrices is given in the following 

\smallskip
\noindent
\textbf{Theorem A2.}
\emph{
Let $X \in \bC^{n \times n}$ be an invertible 
matrix and $V, W \in \bC^{n \times k}$ be 
arbitrary matrices, then we have
\be
\det(X + V W^*) 
= \det(X) \det(\1_k + W^* X^{-1} V).
\ee 
}

\medskip
\noindent
In particular, if $X$ is perturbed by a multiple
of the elementary matrix $e_{a, b}$, then we can 
write
\be
\det(X + \alpha e_{a, b})
= \det(X) + \alpha \cC_{a, b},
\label{rank_1_formula}
\ee
where $\cC_{a, b}$ is the cofactor of $X$ 
associated with entry $X_{a, b}$, i.e., it 
is $(-1)^{a + b}$ times the 
$(n - 1) \times (n - 1)$ minor obtained by 
deleting the $a$th row and the $b$th column 
of $X$. For analogue rank two perturbations 
we have 
\be
\det(X + \alpha e_{a, b} + \beta e_{c, d}) 
= \det(X) + \alpha \cC_{a, b}
+ \beta \cC_{c, d}
+ \alpha \beta 
(\cC_{a, b} \cC_{c, d} - \cC_{a, d} \cC_{c, b}) 
\det(X)^{-1}.
\ee
Finally, if $X$ is an invertible Hermitian matrix,
the above formula simplifies to
\be
\det(X + \alpha e_{a, b} + \overline{\alpha} e_{b, a}) 
=  \det(X) 
+ \alpha \cC_{a, b} 
+ \overline{\alpha} \overline{\cC}_{a, b} 
+ \vert \alpha \vert^2
(\vert \cC_{a, b} \vert^2 - \cC_{a, a} \cC_{b, b}) 
\det(X)^{-1}.
\label{rank_2_formula}
\ee
Since any quadratic submatrix of a Cauchy 
matrix is Cauchy again, the above determinant 
formulae offer a relatively painless 
way to compute the determinants of perturbed 
Cauchy matrices.

\medskip
\noindent
\textbf{Acknowledgments.}
This work was partially supported by the Hungarian
Scientific Research Fund (OTKA) under grant
K 77400.
We thank L. Feh\'er (Univ. Szeged) for useful 
comments on the draft of the paper.

\end{document}